\documentclass[journal]{IEEEtran}

\ifCLASSINFOpdf
\else
   \usepackage[dvips]{graphicx}
\fi
\usepackage{url}

\usepackage{graphicx}

\usepackage{caption} 
\usepackage{tabularx}
\usepackage{float}
\usepackage{array}  
\usepackage{booktabs} 
\usepackage{multirow}
\usepackage{makecell}

\usepackage{soul} 
\usepackage{color, xcolor} 

\usepackage{bbding}

\usepackage{amsmath}

\usepackage{amsfonts,amssymb} 
\usepackage{bbding}  
\usepackage{caption} 
\usepackage{tabularx} 
\usepackage{float}
\usepackage{array}  
\usepackage{booktabs} 
\usepackage{multirow}
\usepackage{makecell} 
\usepackage{color}
 
\usepackage{subfigure} 

\newcommand{\mybluehl}[1]{\textcolor{black}{#1}} 

\begin{document}

\title{\mybluehl{Audio Event-Relational Graph Representation Learning for Acoustic Scene Classification}}

\author{Yuanbo Hou, \IEEEmembership{Student Member, IEEE}, Siyang Song, Chuang Yu, Wenwu Wang, \IEEEmembership{Senior Member, IEEE},\\ Dick Botteldooren, \IEEEmembership{Senior Member, IEEE}
\thanks{Y. Hou and D. Botteldooren are with the WAVES Research Group, Ghent University, Belgium (e-mail: \{Yuanbo.Hou, Dick.Botteldooren\}@UGent.be).}

\thanks{S. Song is with the School of Computing and Mathematical Science, University of Leicester, UK (e-mail: ss1535@leicester.ac.uk).}
\thanks{C. Yu is with the UCL Interaction Centre, University College London, UK (e-mail: chuang.yu@ucl.ac.uk).}
\thanks{W. Wang is with the Centre for Vision, Speech, and Signal Processing, University of Surrey, Guildford GU27XH, UK (e-mail: w.wang@surrey.ac.uk).}

}

\maketitle

\begin{abstract} 
Most deep learning-based acoustic scene classification (ASC) approaches identify scenes based on acoustic features converted from audio clips containing mixed information entangled by polyphonic audio events (AEs). 
However, these approaches have difficulties in explaining what cues they use to identify scenes.
This paper conducts the first study on disclosing the relationship between real-life acoustic scenes and semantic embeddings from the most relevant AEs. 
\mybluehl{Specifically, we propose an \textbf{e}vent-\textbf{r}elational \textbf{g}raph representation \textbf{l}earning (ERGL) framework for ASC to classify scenes, and simultaneously answer clearly and straightly which cues are used in classifying.
In the event-relational graph, embeddings of each event are treated as nodes, while relationship cues derived from each pair of nodes are described by multi-dimensional edge features.
} Experiments on a real-life ASC dataset show that the proposed ERGL achieves competitive performance on ASC by learning embeddings of only a limited number of AEs.
The results show the feasibility of recognizing diverse acoustic scenes based on the audio event-relational graph.
Visualizations of graph representations learned by ERGL are available here
{{(\textcolor{blue}{\underline{https://github.com/Yuanbo2020/ERGL}})}}.
\end{abstract}

\vspace{-0.1cm}

\IEEEpeerreviewmaketitle

\section{Introduction}

\IEEEPARstart{A}coustic scene classification (ASC) aims to classify an audio clip into a pre-defined semantic category, indicating the acoustic environment where the clip is captured (e.g., park, mall, or bus) \cite{acoustic_scene}. 
ASC provides a broad description of the acoustic environment, which can assist intelligent agents in quickly understanding their surrounding environment. As a result, it is useful for various applications, such as 
\mybluehl{
sound source recognition 
\cite{sound_source}\cite{d_case}\cite{pham2023lightweight}\cite{chime}\cite{houjoint}, well-being assistance \cite{well-being}\cite{kim2018sound}\cite{well_being}, and audio-visual scene recognition \cite{dcase_2016}\cite{mmsp}\cite{wang2021audio}\cite{barchiesi2015acoustic}\cite{audiovisual_scnee}.}

Typical deep learning-based ASC methods consist of three steps: 1) Converting a time-domain audio signal to a time-frequency spectrogram, which is used as input acoustic features;
2) Feeding these features to neural networks to obtain high-level representations;
3) Recognizing acoustic scenes based on such high-level representations.  
For example, Ren et al. \cite{Ren2018} utilize a CNN-based model with mel features, where attention-based pooling layers are used to reduce the dimension of representations. 
The spatial pyramid pooling approach is used by CNN in \cite{basbug2019acoustic} to provide various resolutions for ASC. 
\mybluehl{
Apart from mel-based features, wavelet-based deep scattering spectrum \cite{li_icmew} is introduced for ASC.
To explore the instance-level information of audio clips, multiple-instance learning \cite{choi2022instance} is used in ASC.
The higher-order temporal information of acoustic features is exploited by convolutional recurrent neural networks (CRNN) with bidirectional recurrent layers \cite{GAP} and with spatio-temporal attention pooling \cite{phan2019spatio}.}
In addition, attentional graph convolutional networks are used for audio-visual scene classification \cite{attention_graph}.
Given the intrinsic relationship between acoustic scenes and audio events (AEs), some studies jointly analyze scenes and events based on multi-task learning (MTL) \cite{Bear2019TowardsJS}\cite{tonami2021joint}.
\mybluehl{
Relation-guided ASC \cite{RGASC} is proposed to exploit the implicit relations between coarse-grained scenes and fine-grained AEs.}

\label{ssec:figure-f}
\begin{figure*}[t] 
	\setlength{\abovecaptionskip}{0cm}  
	\setlength{\belowcaptionskip}{-0.4cm}   
	\centerline{\includegraphics[width = 0.85 \textwidth]{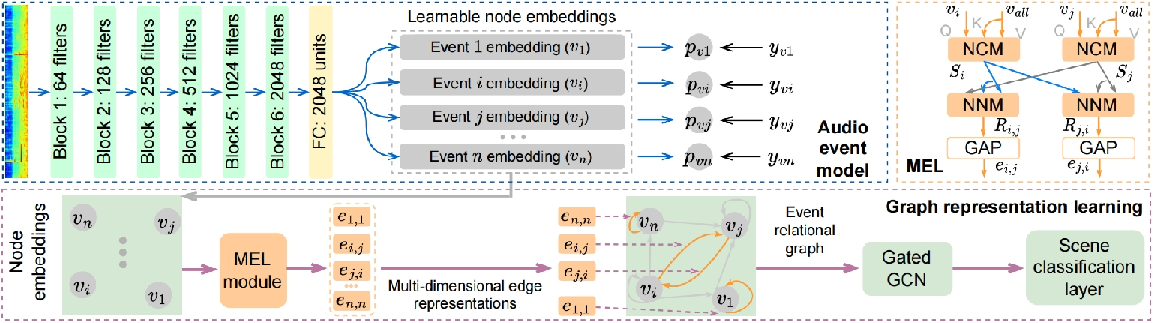}}
	\caption{The framework of the proposed scene-dependent \textbf{e}vent-\textbf{r}elational \textbf{g}raph \textbf{l}earning (ERGL) for ASC.}
	\label{model}
\end{figure*}

\mybluehl{The features used in the methods above often contain polyphonic AEs information.}
These features capture both useful and irrelevant information, as well as noise, which are then used for ASC by recognition systems.
\mybluehl{However, it is difficult to explain what cues in audio clips are used by these approaches to recognize acoustic scenes (ASs).} 
In real life, it is natural for humans to recognize ASs based on the semantically meaningful AEs contained in them, despite variations in relations among the occurring AEs in ASs \cite{dick}. 
\mybluehl{
This paper proposes audio event-relational graph representation learning (ERGL) to classify scenes and simultaneously clearly answer which cues are used in classifying.  
Inspired by the multi-dimensional edge learning in graph-based image analysis \cite{song2021learning}\cite{luo2022learning}, we introduce it into the proposed audio-based ERGL to enhance scene-dependent semantic relationships between non-graph AEs in end-to-end training.
}
These scene-dependent event-relational graphs (ERGs) contain not only the activation of AEs (i.e. nodes in the graph) in the audio clip, but also their relations (represented as edges) that are relevant to ASC tasks.
\mybluehl{
The graph in ERGs is a single graph. Thus, spatial-temporal graph neural networks  \cite{spatio_temporal_graph}\cite{st_gnn}, which aim to capture spatial and temporal dependencies of graph sequences or multigraphs, are unsuitable for modelling ERGs. ERGs are fed to a gated graph convolutional network (Gated GCN) \cite{gated_GCN_2016} for ASC.} 
Experiments show that graph representations learned by ERGL, from only several explicit audio event semantic embeddings, can facilitate the discrimination between different ASs.


\section{Audio event-relational graph learning}
\label{sec:format}

\mybluehl{
This section discusses the proposed approach for learning the scene-related event-relational graph (ERG) representation from non-graph audio clips in an end-to-end manner.
}
First, we learn a set of AEs embeddings, where each embedding ($v_i$) contains the $i$-th audio event-related information, and is treated as the $i$-th node in the ERG. 
\mybluehl{
Then, we learn a pair of multi-dimensional edge features ($e_{i,j}$, $e_{j,i}$) to describe the relations between each pair of nodes ($v_i$, $v_j$).
Thus, the obtained graph explicitly describes the occurrence of a set of AEs and their relations relevant to the given scene.  
Finally, the obtained ERG with $n$ nodes and $n\times n$ edges is fed into Gated GCN for ASC.}

\vspace{-0.1cm}
\subsection{Audio event node embedding learning}\label{sec:node_event_label}

\mybluehl{We first propose a model derived from PANNs \cite{kong2020panns} for node embedding generation, where each describes a specific AE.} 
As shown in Fig.~\ref{model}, the spectrogram of the audio clip is fed into a set of convolutional blocks. Each block contains two convolutional layers with kernels of size 3 × 3, a batch normalization \cite{batchnormal}, and a ReLU function \cite{relu}. Then, a fully-connected (FC) layer with 2048 units generates joint representations for all AEs. 
\mybluehl{Unlike PANNs, we employ $n$ independent FC layers with 64 units to learn $n$ embeddings separately, where each embedding describes a unique pre-defined audio event.}

During training, each event embedding $v_i$ is fed into the following event classification layer to predict the corresponding event probability $p_{vi}$, \mybluehl{where mean squared error (MSE) loss is used to measure the distance between the prediction $p_{vi}$ and the label $y_{vi}$ (i.e., $\mathcal{L}_{\text{event}} = MSE(p_{vi}, y_{vi})$).  
To train the audio event node embedding generation module, we employ PANNs, which contains 527 classes of AEs, to generate pseudo labels of AEs.} 
This produces a $527$-dimensional soft pseudo label $y = [y_{v1}, y_{v2}, ..., y_{v{527}}]$ for each audio clip, describing the occurrence probabilities of 527 classes of AEs. 
Since real-world acoustic scene datasets rarely have all 527 classes of AEs, i.e., the number of occurred AEs would be much smaller than 527, we rank all AEs by accumulating their probabilities in all training data, and use a set of top-ranked (Top $n$) AEs with the highest overall probability describing each scene. \mybluehl{As a result, each graph contains $n$ nodes and $n \times n$ edges.}

\vspace{-0.1cm}
\subsection{Audio event-relational edge feature learning}

Once all audio event (node) embeddings are obtained, we propose a \textbf{m}ulti-dimensional \textbf{e}dge feature \textbf{l}earning (MEL) module to learn scene-related relations between each pair of AEs.
\mybluehl{Here, our hypothesis is that the co-occurrence patterns of all event pairs may include key clues for ASC tasks.} 
The proposed MEL module in Fig.~\ref{model} consists of two sub-modules, the \text{node-context relation modelling (NCM)} and the \text{node-node relation modelling (NNM)}. NCM first learns the ASC-task-specific relation cues between each node (event) and the global context (scene), \mybluehl{generating a scene-aware representation to represent each node.} Then, \mybluehl{NNM models the semantic relations between nodes in each node pair, to generate the final multi-dimensional edge feature describing the AE-based scene-aware relations between each pair of nodes.}

\textbf{NCM.} For each node $v_i$, NCM conducts cross-attention \cite{Transformer} between it and the global contextual representation $v_{\text{all}}$ consisting of the mean of all node features, where node $v_i$ is used as the query, and $v_{all}$ is employed as the key and value.
\begin{equation}
\setlength{\abovedisplayskip}{1.5pt}
\setlength{\belowdisplayskip}{1.5pt} 
\text{NCM}(\mathbf{Q, K} )=\Phi(\mathbf{{QW}}_q\mathbf{(KW}_k)^T/\sqrt{d_{k}})\mathbf{KW}_v 
\label{self-attention}
\end{equation}   
where $\Phi$ is the softmax function, $\mathbf{W}_{\{q, k, v\}}$ are learnable weights, and $d_k$ is a factor equal to the number of channels in $\mathbf{K}$. As a result, the obtained representations $\mathcal{S}_{i}$ ($i = 1, 2, \cdots, n$) encode scene-aware cues for each audio event. 
\begin{equation}
\setlength{\abovedisplayskip}{1.5pt}
\setlength{\belowdisplayskip}{1.5pt}  
\mathcal{S}_{i}=\text{NCM}(v_i, v_{\text{all}}), \quad \mathcal{S}_{j}=\text{NCM}(v_j, v_{\text{all}})
\end{equation}

\textbf{NNM.} \mybluehl{After extracting all scene-aware event node features, NNM module then models the semantic relationship between nodes by capturing multi-dimensional (m-d) edge features.} In particular, NNM consists of cross-attention \cite{Transformer} and global average pooling (GAP) \cite{GAP} layer, \mybluehl{which takes a pair of scene-aware event features ($\mathcal{S}_{i}$, $\mathcal{S}_{j}$) as input, and a pair of m-d edge features ($e_{i, j}$,  $e_{j, i}$) as output. In detail, NNM first conducts:} 
\begin{equation}
\setlength{\abovedisplayskip}{1.5pt}
\setlength{\belowdisplayskip}{1.5pt}  
\mathcal{R}_{i, j}=\text{NNM}(\mathcal{S}_{j}, \mathcal{S}_{i}), \quad \mathcal{R}_{j, i}=\text{NNM}(\mathcal{S}_{i}, \mathcal{S}_{j})
\end{equation} 
where the edge feature $\mathcal{R}_{i, j}$ encodes $\mathcal{S}_{j}$-related cues in $\mathcal{S}_{i}$, and correspondingly, $\mathcal{R}_{j, i}$ encodes $\mathcal{S}_{i}$-related cues in $\mathcal{S}_{j}$. Next, edge features $\mathcal{R}_{i, j}$ and $\mathcal{R}_{j, i}$ are fed into the GAP layer to obtain the multi-dimensional edge feature vectors $e_{i, j}$ and $e_{j, i}$.
\begin{equation}
\setlength{\abovedisplayskip}{1.5pt}
\setlength{\belowdisplayskip}{1.5pt} 
e_{i, j} = \text{GAP}(\mathcal{R}_{i, j}), \quad
e_{j, i} = \text{GAP}(\mathcal{R}_{j, i})
\end{equation} 
\mybluehl{Consequently, the produced $e_{i, j}$ and $e_{j, i}$ capture multiple ASC-task-specific cues related to both event nodes $v_i$ and $v_j$.}

\vspace{-0.2cm}
\subsection{\mybluehl{Scene-aware event-relational graph}}
 
Once the ERG (denoted as $G^0$) that contains $n$ node embeddings $v = \{v_1, v_2, \cdots, v_n \}$ and $n\times n$ multi-dimensional directed edge representations $e = \{e_{1,1}, \cdots, e_{i,j}, \cdots, e_{n,n} \}$ is obtained, \mybluehl{we feed $G^0$ to the Gated GCN \cite{gated_gcn}\cite{gated_GCN_2016} for ASC.}

Since the model contains $U$ GCN layers, its output is $G^U=(v^U,e^U)$, which is a graph with the same topology as $G^0$.  
The $i$-th node represents the activation state of the $i$-th event in the scene. The latent node features in $G^U$ are concatenated as the scene representation and input to the final scene classification layer.
Cross entropy (CE) \cite{celoss} is used as the loss function in ASC between the prediction $p_{s}$ and the scene true label ${y}_{s}$,
$\mathcal{L}_\text{scene} = CE(p_{s}, y_{s})$. 
\mybluehl{
Hence, the final loss of ERGL is $\mathcal{L} = \lambda_1\mathcal{L}_\text{event} + \lambda_2\mathcal{L}_\text{scene}$, where $\lambda_i$ ($i = 1,2$) 
default to 1 in this paper.}
The effect of $U$, which defaults to 2, on the model performance will be explored in the experiments later.

\vspace{-0.2cm}
\section{Experiments and results}
\label{sec:experiments}

\subsection{Dataset, baseline, experimental  setup, and metric}

This paper uses TUT Urban Acoustic Scenes 2018 dataset (UAS) \cite{DCASE2018} with 8640 10-second clips. 
\mybluehl{The UAS contains 10 classes of real-life acoustic scenes,} 24 hours in total.  
However, UAS does not provide labels for audio events.
To obtain the event labels used in Sec.~\ref{sec:result}, pre-trained model PANNs are used to annotate audio clips with 527 classes of AEs pseudo labels.
To compare with other methods on the same test set, we follow the setup of \cite{DCASE2018}, \mybluehl{where the test, training, and validation sets contain 2518, 5509, and 613 samples, respectively.}

In addition to a typical CNN-based approach \cite{DCASE2018} for ASC, this paper also employs attention-based \cite{Ren2018}, \mybluehl{spatio-temporal attention \cite{phan2019spatio}, multiple-instance learning \cite{choi2022instance}, and scene-event joint learning  \cite{Bear2019TowardsJS}\cite{tonami2021joint}\cite{komatsu2020scene}\cite{RGASC} as baselines for comparison. }

Following \cite{kong2020panns}, the log mel spectrogram with 64 bins is used as the acoustic feature, which is extracted by the Short-Time Fourier Transform with a Hamming window of size 1024 and a hop size of 320 samples. 
A batch size of 64 and AdamW optimizer \cite{adamw} with a learning rate of 1e-4 are used to minimize the loss. 
The systems are trained for 400 epochs.
The accuracy (\textit{Acc}) \cite{acoustic_scene} is used as the performance metric.

\vspace{-0.2cm}
\subsection{Results and analysis}\label{sec:result}

The pseudo labels composed of probability outputs by PANNs \cite{kong2020panns} are used as supervision information for training the audio event model, as shown in Fig.~\ref{model}. 
The accuracy of pseudo labels that do not involve human verification cannot be evaluated. 
\mybluehl{Since our goal is ASC, we will mainly show ASC results, rather than the accuracy of pseudo labels on AEs.}


		
		

\begin{table}[b]\footnotesize 
	\setlength{\abovecaptionskip}{0.1cm}   
	\setlength{\belowcaptionskip}{-0.2cm}   
	\renewcommand\tabcolsep{0.8pt} 
	\centering 
\caption{\small{\textit{Acc} of ERGL at different $n$ values on the validation set.}}
	\begin{tabular}
	{p{1cm}<{\centering}|
    p{1.2cm}<{\centering}
	p{2cm}<{\centering}| 
	p{1cm}<{\centering}|
	p{1.2cm}<{\centering}
	p{2cm}<{\centering}} 

\hline
		{\#} & Top $n$ & $Acc$  (\%) & {\#} & Top $n$ & $Acc$  (\%)  \\ 
		
		\specialrule{0em}{0.1pt}{0.1pt}
		
        \hline
        \specialrule{0em}{0.1pt}{0.1pt} 
		1 & 10 &  94.03 $\pm$ 2.41  & 5 & 100 &  94.22 $\pm$ 2.91   \\ 
  2 & 25 &  \textbf{95.92} $\pm$ 2.29  & 6 & 200 &  93.05 $\pm$ 2.55 \\ 
  3 & 50 &  94.65 $\pm$ 2.75   & 7 & 300 &  92.46 $\pm$ 3.89 \\ 
  4 & 75 &  94.47 $\pm$ 1.89  & 8 &  400 &  92.29 $\pm$ 2.07 \\  
		\hline
	\end{tabular}
	\label{tab:topn}
\end{table}

\textbf{The number $n$ of audio events.} 
We first evaluate the impact of the choice of $n$, i.e.,  the number of top AEs used in ERGL, on the performance of ERGL.
\mybluehl{The \textit{Acc} of ERGL does not increase monotonically as $n$ increases, as shown in Table~\ref{tab:topn}.} 
The reason may be that as the number of events $n$ increases, the number of nodes in the graph increases linearly, but the number of edges in the graph grows in the order of $n^2$, which sharply increases the burden for learning the multi-dimensional edge features with the MEL module.     
The increased number of parameters does not provide more useful information to the model, but may compromise its performance.

The ERGL works best when $n=25$, indicating that only using 25 classes of AEs can describe the 10 classes of scenes in the dataset.
In preprocessing the distribution of AEs in each scene, we also found that the 25 classes of AEs automatically selected by the model cover most of the dominant AEs in each scene. Therefore, we set $n=25$ in the following experiments.

\textbf{The number ($U$) of GCN layers.} 
Table~\ref{tab:layers} explores the ERGL performance under different numbers of GCN layers.
\mybluehl{The results in Table~\ref{tab:layers} illustrate that increasing $U$ does not lead to better results.}
The reason for this may be that the 2-layer Gated GCN already achieves a good balance between model performance and computational efficiency on the graph consisting of semantic embeddings of 25 classes of AEs, and also that adding extra layers would make the model deeper and harder to train. \mybluehl{Subsequent experiments will set $U$ as 2.}


  

\begin{table}[H]\footnotesize 
	\setlength{\abovecaptionskip}{0.1cm}  
	\setlength{\belowcaptionskip}{-0.1cm}  
 \setlength{\belowcaptionskip}{-0cm}  
	\renewcommand\tabcolsep{0.8pt} 
	\centering 
\caption{\small{\textit{Acc} of ERGL at different $U$ layers on the validation set.}}
	\begin{tabular}
	{p{1cm}<{\centering}|
    p{1.2cm}<{\centering}
	p{2cm}<{\centering}| 
	p{1cm}<{\centering}|
	p{1.2cm}<{\centering}
	p{2cm}<{\centering}} 

\hline
		{\#} & $U$ & $Acc$  (\%) & {\#} & $U$ & $Acc$  (\%)  \\  
		\specialrule{0em}{0.1pt}{0.1pt} 
        \hline
        \specialrule{0em}{0.1pt}{0.1pt} 
		1 & 1 &  94.42 $\pm$ 1.97  & 4 & 4 &  95.09 $\pm$ 2.74    \\ 
  2 & 2 &  \textbf{95.92} $\pm$ 2.29  & 5 & 5 & 94.79 $\pm$ 2.43  \\ 
  3 & 3 &  95.66 $\pm$ 2.58  & 6 & 6 &  94.29 $\pm$ 1.92\\     
		\hline
	\end{tabular}
	\label{tab:layers}
\end{table}

\vspace{-0.1cm}
\mybluehl{
\textbf{Ablation study of AE-based relations in ERGL.} 
The MEL module in ERGL aids in capturing multi-dimensional AE-based semantic relations between nodes. In MEL, NCM learns the relation between the node and global contextual representation to represent the node, while NNM uses cross-attention to capture the semantic relations between nodes.
To investigate how well NNM captures semantic relations,
Table~\ref{tab:ablation_study} presents the ablation study to compare the performance of ERGL with (w/) and without (w/o) AE-based relations.}

\begin{table}[H]\footnotesize 
	\setlength{\abovecaptionskip}{0.1cm} 
	\setlength{\belowcaptionskip}{-0.1cm}  
 \setlength{\belowcaptionskip}{-0cm}  
	\renewcommand\tabcolsep{0.8pt} 
	\centering 
\caption{\small{\mybluehl{Ablation study of AE relation-related MEL in ERGL.}}}
	\begin{tabular}
	{p{1cm}<{\centering}|
 p{1cm}<{\centering}|
    p{1.6cm}<{\centering}|
	p{1.6cm}<{\centering}|
    p{1.6cm}<{\centering}|
    p{1.6cm}<{\centering}} 
\hline
	\mybluehl{\multirow{2}{*}{\makecell[c]{MEL}}} & \mybluehl{NCM} & \mybluehl{\XSolidBrush}  & \mybluehl{\CheckmarkBold} & \mybluehl{\XSolidBrush} & \mybluehl{\CheckmarkBold} \\  
        \cline{2-6}  
    & \mybluehl{NNM} & \mybluehl{\XSolidBrush} & \mybluehl{\XSolidBrush} & \mybluehl{\CheckmarkBold} & \mybluehl{\CheckmarkBold} \\
    \hline  
     \multicolumn{2}{c|}{Test set $Acc$ (\%)} &   
    \mybluehl{73.35}$\pm$\mybluehl{1.98} & 
    \mybluehl{74.42}$\pm$\mybluehl{1.99} & 
    \mybluehl{75.69}$\pm$\mybluehl{1.54} & \mybluehl{\textbf{78.08}}$\pm$\mybluehl{2.06} \\

  
		\hline
	\end{tabular}
	\label{tab:ablation_study}
\end{table}

\begin{figure}[t] 
	\setlength{\abovecaptionskip}{0.1cm}  
	\setlength{\belowcaptionskip}{-0.4cm}   
	\centerline{\includegraphics[width = 0.5 \textwidth]{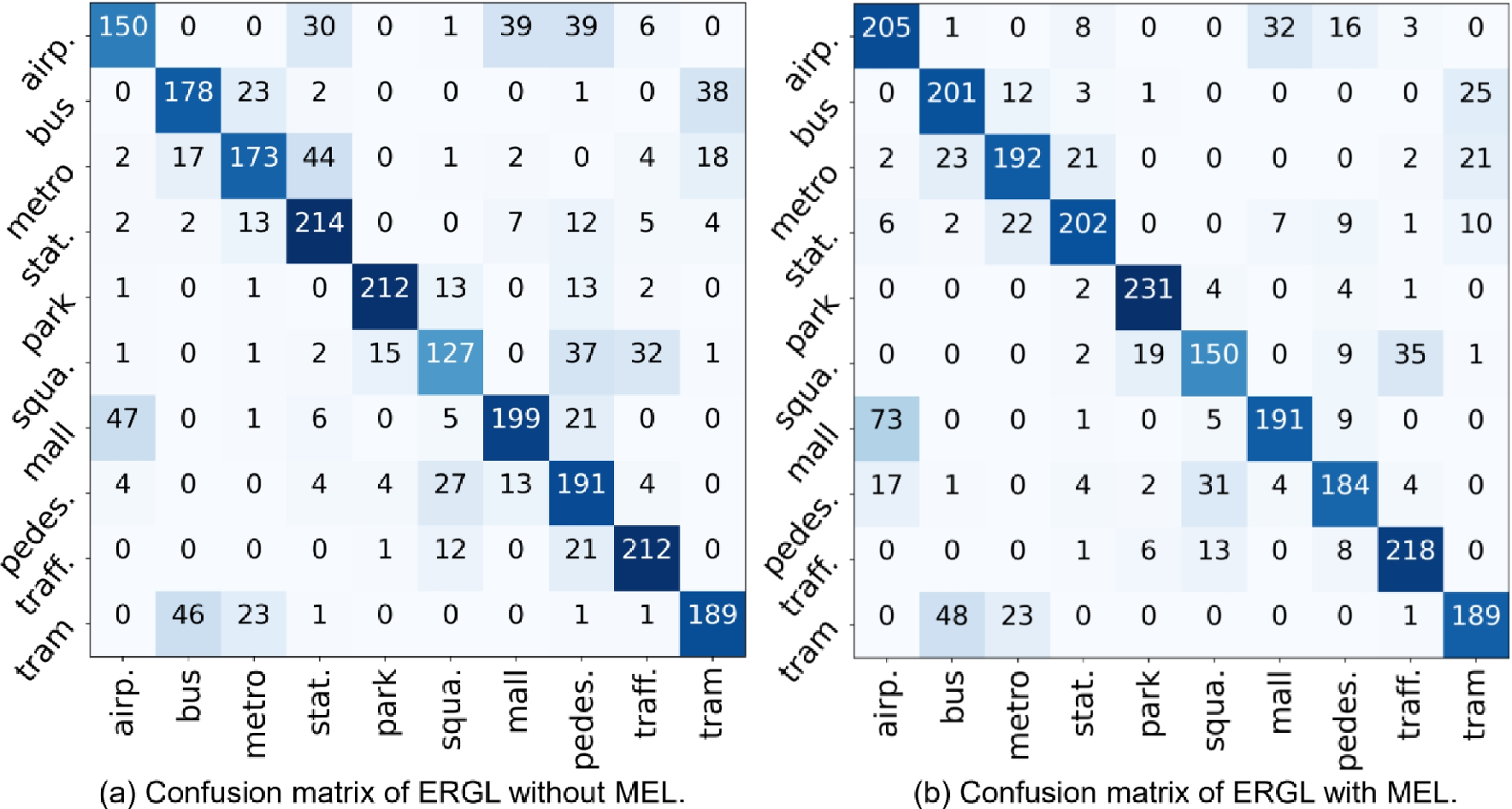}}
	\caption{\small{\mybluehl{
 Confusion matrix of ERGL w/o and w/ AE-based relational edges on the test set. (X-axis: Predicted label; Y-axis: True label.)}}}
	\label{confusion_matrix_tsne}
\end{figure}

\vspace{-0.1cm}
\mybluehl{
Table~\ref{tab:ablation_study} shows that ERGL w/ NNM outperforms ERGL w/ NCM. For graph representation learning in ERGL, NNM, which aims to capture semantic relations between nodes, is more valuable than NCM, which focuses on learning relations between the node and global contextual representation. 
This shows that NNM based on cross-attention capturing AE-based semantic relations is effective.
Fig.~\ref{confusion_matrix_tsne} shows confusion matrices of ERGL w/o and w/ AE-based relations to explore where the AE-based relational approach works and where it does not. 
}

\label{ssec:figure-f}

\mybluehl{
Fig.~\ref{confusion_matrix_tsne} illustrates that AE-based relational edges effectively help ERGL improve its accuracy in 6 scenes:  
``\textit{airp.} (airport), \textit{bus}, \textit{metro}, \textit{park}, \textit{squa.} (public square), \textit{traff.} (street traffic)".
The accuracy of  \textit{airp.} is improved the most, mainly because the misclassified samples between \textit{stat.} (metro station), \textit{pedes.} (street pedestrian), and \textit{airp.} is reduced from 30 and 39 to 8 and 16, respectively. 
In contrast, introducing MEL increases the misclassified samples of \textit{stat.} and \textit{metro}. Even for humans, it is challenging to distinguish these similar scenes relying on audio only.
In short, introducing AE-based relational edges can effectively improve the performance of ERGL in 6 scenes, increasing its \textit{Acc} from 73.35\% to 78.08\% in Table~\ref{tab:ablation_study}.}

\textbf{Comparison with non-ensemble ASC methods.}
\mybluehl{Table \ref{tab:asc_model} shows the results of models on the same test set.}
In the fixed mode~\cite{interspeech2020hyb}, the parameters of PANNs are not updated in training.
The result of fixed-mode PANNs is comparable to  Baseline, implying that PANNs with AEs knowledge, which is learned from AudioSet \cite{aduioset}, have a certain discriminative ability for scenes. 
In contrast, ERGL using just audio event embeddings improves ASC accuracy even though these event embeddings are learned from pseudo labels without verification.
The proposed end-to-end EGRL without data augmentations offers competitive results.
\mybluehl{This illustrates that ERGL can effectively discriminate different scenes by relying only on several scene-aware events semantic embeddings.}

\begin{table}[H] \footnotesize 
 \setlength{\abovecaptionskip}{0.1cm}  
	\setlength{\belowcaptionskip}{-0.1cm}
	\renewcommand\tabcolsep{1pt} 
	\centering
	\caption{\small{Comparison of non-ensemble systems on the test set.}}
	\begin{tabular}{
	p{3.6cm}<{\centering}|
	p{3.9cm}<{\centering}|
	p{1.0cm}<{\centering}
	} 
	    \hline 
		System & Model structure & \textsl{Acc (\%)}\\ 
		\hline
		PANNs \cite{kong2020panns} (Fixed mode) & VGG-like CNN & 56.9 \\ 	
		Baseline \cite{DCASE2018} & CNN & 59.7 \\  
  
		NNF-CNNEns \cite{Nguyen2018a} & CNN and nearest neighbor filters & 69.3 \\  

\mybluehl{Spatio-temporal Attention  \cite{phan2019spatio}} & \mybluehl{CRNN} & \mybluehl{72.5} \\ 

  \mybluehl{Attention-based CNN \cite{Ren2018}} & Attention-Based CNN & 72.6 \\ 
		PANNs (Fine-tuning mode) & VGG-like CNN & 73.8 \\ 

  \mybluehl{Instance-based ASC \cite{choi2022instance}} & \mybluehl{CNN} & \mybluehl{73.9} \\ 
  
  Wavelet-based  spectrum \cite{li_icmew} & CRNN & 76.6 \\
  Proposed ERGL & CNN and Graph Learning & \textbf{78.1} \\ 
	\hline
	\end{tabular}
	\label{tab:asc_model}
\end{table}

\vspace{-0.1cm}
\textbf{Comparison with scene-event joint methods.}
The ERGL infers target scenes based on the ERG in the scene, which is scene-event joint analysis. Table \ref{tab:asc_joint} compares ERGL with other scene-event joint methods.
\mybluehl{The model using the same latent space to classify scenes and events \cite{Bear2019TowardsJS} performs the worst.}
The reason may be that real-life scenes and AEs differ at the semantic level and the feature space.
Papers \cite{tonami2021joint}\cite{komatsu2020scene} use shared base and separated high-level features to identify scenes and AEs.
RGASC \cite{RGASC} exploits the scene-event relationship to guide the model to achieve mutually beneficial scene-event classification.
\mybluehl{The ERGL relies only on semantic embeddings of AEs to achieve one-way event-to-scene inference, and recognizes scenes based on the corresponding explicit semantic ERG.
Notably, ERGL, which only needs 25 classes of AEs' information, outperforms RGASC with 527 classes of AEs' information.}
Overall, the ERGL achieves promising results, demonstrating the feasibility of ASC based on the ERG.

\begin{table}[b]\footnotesize
\setlength{\abovecaptionskip}{0.1cm}  
\setlength{\belowcaptionskip}{-0.2cm}
\renewcommand\tabcolsep{1pt} 
	\centering
	\caption{\small{\textit{Acc} of scene-event joint analysis methods on test set.}}
	\begin{tabular}{
	p{0.7cm}<{\centering}|
	p{6.cm}<{\centering}|
	p{1.4cm}<{\centering}
	}
	    \hline
		\# & Method &  \textsl{Acc} (\%)\\
		\hline
		1 & Scene and event jointly classification \cite{Bear2019TowardsJS} &   52.35 \\ 
		
		2 & MTL-based event and scene analysis   \cite{tonami2021joint} &    61.69 \\
		
		3 & Conditional scene and event recognition \cite{komatsu2020scene} &   66.39 \\ 
		
		4 & Relation-guided ASC (RGASC) \cite{RGASC} &  77.35 \\

  5 & The proposed ERGL &  \textbf{78.08} \\
	
	\hline
	\end{tabular}
	\label{tab:asc_joint}
\end{table}

\begin{figure}[t] 
	\setlength{\abovecaptionskip}{0.1cm}  
	\setlength{\belowcaptionskip}{-0.2cm}   
	\centerline{\includegraphics[width = 0.5 \textwidth]{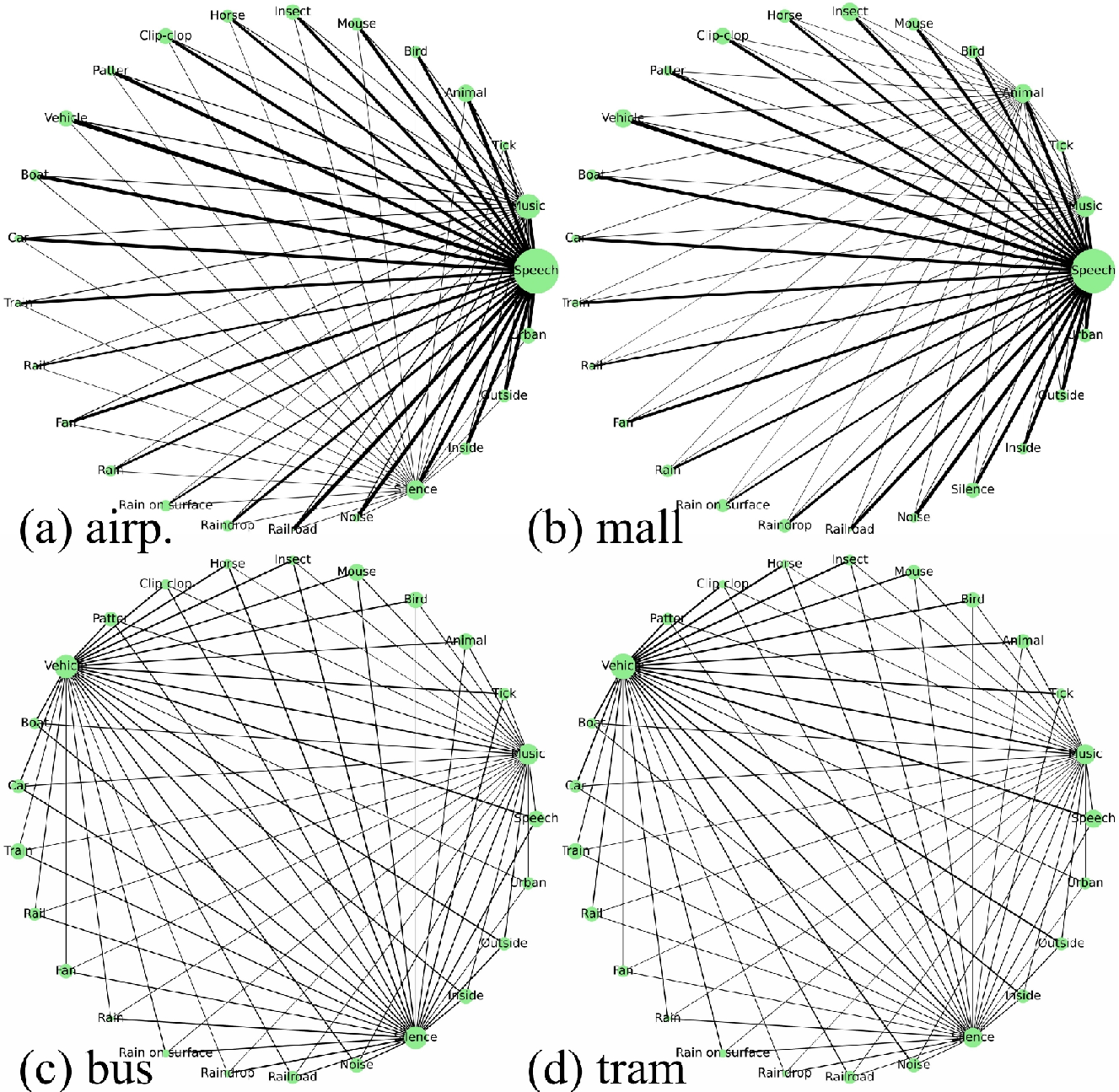}}
	\caption{\small{\mybluehl{Graph structures of test audio samples from different scenes. 
 Green dots represent  25 classes of AEs used in this paper. 
A larger dot denotes a higher probability of the event. 
A thicker line denotes a larger edge value between nodes in the graph representation.
(\textit{For visualization, a node is considered inactive if its probability is less than 0.1. 
Edges between two inactive nodes are not displayed. 
Multi-dimensional features of the edge are represented by their mean value.})}}}
	\label{4_scene_graph_show}
\end{figure}

\textbf{Analysis of confusion.}
To further explore the reasons for the misclassification between scenes for graph representation-based classification. 
Fig.~\ref{4_scene_graph_show} presents the structure of graph representations of audio clips from different scenes.
In Fig.~\ref{4_scene_graph_show} (a) and (b), the dominant AEs in \textit{airp.} and \textit{mall} scenes are \textit{speech} and \textit{music}, so the connections in the graph are mainly gathered around \textit{speech} and \textit{music}. 
This may be the reason that ERGL confuses the \textit{airp.} and \textit{mall} scenes in Fig.~\ref{confusion_matrix_tsne}. In addition to these similarities, the third focused audio event in Fig.~\ref{4_scene_graph_show} (a) is \textit{silence}, while that in Fig.~\ref{4_scene_graph_show} (b) is \textit{animal} sounds, which reflects the differences between the two scenes.
In Fig.~\ref{4_scene_graph_show} (c) and (d), the dominant AEs in \textit{bus} and \textit{tram} scenes are \textit{vehicle}, \textit{music}, \textit{speech}, \textit{train}, \textit{car} and \textit{silence}.
\mybluehl{And the AEs with dominant connections are the same: \textit{vehicle}, \textit{music} and \textit{silence}.
With similar dominant events and graph structures, the model tends to be confused by these similar scenes.}

\label{ssec:figure-f}

\vspace{-0.2cm}
\section{CONCLUSION}
\label{sec:CONCLUSION}

\mybluehl{To perform ASC and simultaneously clearly answer which cues are used in classifying, we propose a scene-dependent audio event-relational graph representation learning method for ASC, which represents acoustic scenes by a set of scene-aware nodes with explicit AEs semantic embeddings, and specifically produces scene-task-relevant multi-dimensional edge features to describe AE-based semantic relations between nodes.}
Experiments show that ERGL achieves competitive ASC performance by learning ERGs, which are constructed on semantic embeddings of only a limited number of AEs.

 \section{ACKNOWLEDGEMENTS}
\label{sec:ACKNOWLEDGEMENTS}
The WAVES Research Group received funding from the Flemish Government under the “Onderzoeksprogramma Artificiële Intelligentie (AI) Vlaanderen” programme.

\vfill\pagebreak

\label{sec:refs}
 
\bibliographystyle{IEEEbib}
\bibliography{main}

\end{document}